\def\year{2021}\relax
\documentclass[letterpaper]{article} 
\usepackage[switch]{lineno} 
\usepackage{aaai21}  
\usepackage{times}  
\usepackage{helvet} 
\usepackage{courier}  
\usepackage[hyphens]{url}  
\usepackage{graphicx} 
\def\UrlFont{\rm}  
\usepackage{natbib}  
\usepackage{caption} 
\usepackage{makecell}
\usepackage{listings}
\usepackage{mathtools}
\usepackage{algorithm}
\usepackage[algo2e,ruled,vlined]{algorithm2e} %

\title{Focusing Knowledge-based Graph Argument Mining via Topic Modeling}
\author{
    Patrick Abels\textsuperscript{\rm 1},
    Zahra Ahmadi\textsuperscript{\rm 1}, 
    Sophie Burkhardt\textsuperscript{\rm 1}, 
    Benjamin Schiller\textsuperscript{\rm 2}, \\
    Iryna Gurevych\textsuperscript{\rm 2}, 
    Stefan Kramer\textsuperscript{\rm 1}\\
}
\affiliations {
    \textsuperscript{\rm 1} Johannes Gutenberg University Mainz, Germany 
    \\pabels@students.uni-mainz.de, zaahmadi@uni-mainz.de, burkhardt,kramer@informatik.uni-mainz.de\\
    \textsuperscript{\rm 2} Ubiquitous Knowledge Processing Lab, Department of Computer Science, Technical University of Darmstadt 
    \url{www.ukp.tu-darmstadt.de} 
}

\begin{document}

\maketitle

\begin{abstract}
Decision-making usually takes five steps: identifying the problem, collecting data, extracting evidence, identifying pro and con arguments, and making decisions. Focusing on extracting evidence, this paper presents a hybrid model that combines latent Dirichlet allocation and word embeddings to obtain external knowledge from structured and unstructured data. We study the task of sentence-level argument mining, as arguments mostly require some degree of world knowledge to be identified and understood. Given a topic and a sentence, the goal is to classify whether a sentence represents an argument in regard to the topic. We use a topic model to extract topic- and sentence-specific evidence from the structured knowledge base Wikidata, building a graph based on the cosine similarity between the entity word vectors of Wikidata and the vector of the given sentence. Also, we build a second graph based on topic-specific articles found via Google to tackle the general incompleteness of structured knowledge bases. Combining these graphs, we obtain a graph-based model which, as our evaluation shows, successfully capitalizes on both structured and unstructured data. 

\end{abstract}

\section{Introduction}
The emerging field of argument mining aims to gather data with an argumentative context regarding a topic of interest, and by this, support decision making. Following Stab \textit{et al.}~\shortcite{Gurevych:2018}, we define an argument as a combination of a topic and a sentence holding evidence towards this topic (Figure~\ref{gfx:problem_input}). 
However, without accessing some relevant world knowledge, it is difficult to understand the found arguments. 
We tackle this problem by building a local knowledge graph from structured data (Wikidata) and unstructured data (Google search) for each sentence and extracting paths leading from one token to another. We use latent Dirichlet allocation (LDA) \cite{Blei:2003} to improve the quality of connections in the knowledge graph by comparing the context of a corpus, regarding the given sentence and topic, with the properties that each Wikidata entity may be connected with.

\begin{figure}[tb]
    \centering
\begin{lstlisting}[mathescape, frame=single, breaklines, basicstyle=\footnotesize\ttfamily]
Topic: gun control
Sentence: In these days and $\underline{times}$, a lot of us do not feel safe in our own homes or $\underline{offices}$.
\end{lstlisting}
\begin{lstlisting}[mathescape, frame=none, breaklines]
$\textbf{offices}\xrightarrow[]{WIKIFIERED}\textbf{Office (Q12823105)}\xrightarrow[P279]{subclass~of}$
$\textbf{room (Q180516)}\xrightarrow[P279]{subclass~of}\textbf{location (Q17334923)}$
$\xrightarrow[P361]{part~of}\textbf{space (Q107)}\xrightarrow[P361]{part~of}\textbf{\textit{spacetime (Q133327)}}$
$\xrightarrow[P527]{has~part}\textit{Time (Q11471)}\xrightarrow[]{WIKIFIERED}\textit{times}$
\end{lstlisting}
    \caption{A sample argument input (top) and its noisy concept path (bottom).} 
    \label{gfx:problem_input}
\end{figure}

Early work introduced world knowledge into NLP tasks by harnessing manually constructed knowledge bases like Wikidata or DBPedia~\cite{Botschen:2018, Fromm:2019}. Soon after realizing the incompleteness of structured knowledge sources, authors provided world knowledge from unstructured data such as Wikipedia~\cite{Potash:2017}. Having problems with noise in unstructured sources, more recent work combines both structured and unstructured data~\cite{Clark:2019,Lv:2020,Ostendorff:2019}. Motivated by the similarity of argument-based tasks, utilizing and enhancing pre-trained language models like BERT \cite{Devlin:2018} has gained huge popularity \cite{Wang:2020, Xiong:2019, Zhang:2019, Peters:2019, Guan:2020, He:2019, Lauscher:2019, Levine:2019}. In comparison, besides being simpler, our approach focuses more on the knowledge graph itself and how it can be efficiently traversed. Figure \ref{gfx:problem_input} shows a sample argument and how a path is found based on the sentence. We summarize our contributions as follows:
\begin{enumerate}
\item We develop a graph-based approach leveraging evidence from structured
knowledge bases via latent Dirichlet allocation.
\item We propose a comparably efficient dynamic breadth-first search algorithm using word embeddings to create a sparse knowledge graph.
\item We introduce a method to enrich a knowledge graph with unstructured data
from Google via OpenIE.
\item We achieve an average accuracy of 85\% and F1-score of 67\% on the UKP Sentential Argument Mining Corpus.
\end{enumerate}

\begin{figure}[t]
\centering
\includegraphics[width=0.95\columnwidth,trim=8 8 8 8,clip,]{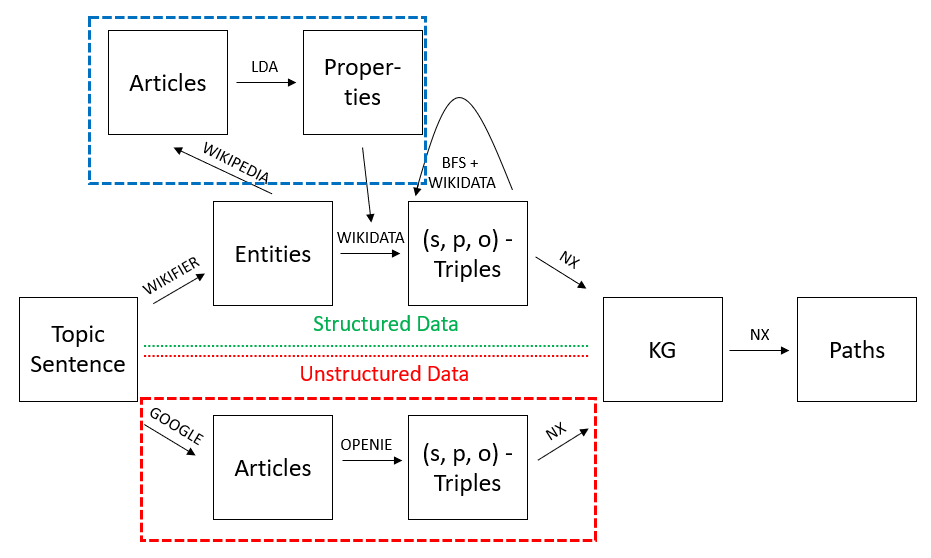}
\caption{The proposed framework: The main core is augmented by LDA-driven property selection (blue box) and unstructured knowledge enrichment via Google search (red box).} \label{gfx:total}
\end{figure}

\section{Related Work}

Recent work that combines both structured and unstructured data proposes new benchmarks on various tasks, including question answering \cite{Clark:2019, Lv:2020}, and document classification \cite{Ostendorff:2019}. Clark \textit{et al.} \shortcite{Clark:2019} solve non-diagram multiple-choice science exams with over 90\% (8th grade) and 83\% (12th grade) accuracy. Their Aristo project consists of three different methods: (1) statistical and information retrieval methods, including searching the exact question in structured datasets, (2) reasoning methods using semi-structured data via OpenIE \cite{Manning:2014}, and (3) large-scale language model methods such as ELMo \cite{Peters:2018} and BERT \cite{Devlin:2018}.

In a similar graph-based approach but on a different problem, Lv \textit{et al.} \shortcite{Lv:2020} build two graphs on structured data from ConceptNet and unstructured data from Wikipedia for the Commonsense Question Answering problem, achieving an accuracy of 75.3\% on the CommonsenseQA dataset \cite{Talmor:2019}. While they select the top 10 sentences regarding the given query as Wikipedia evidence via the Elastic Search engine, we rely on several hundred of topic-specific sentences from Google search ranked via PageRank \cite{Page:1999}. Furthermore, instead of ConceptNet, we use Wikidata as a source of structured data, hence, we use the knowledge from Wikimedia. Additionally, their rules for two nodes of the knowledge graphs being connected are relatively strict: Either one is contained in the other, or they only differ in one word. Focusing on unlabeled data, we do not want two nodes to be this restricted. Instead, we require only one common word between two nodes which adds more flexibility. While they build a graph from structured data with each statement being a node and they topology sort it to avoid cycles, we consider entities as vertices and relations as edges, having no need for sort. With their restrictive search, Lv \textit{et al.} consider 20 nodes in structured data and 10 sentences from unstructured data, while both of our graphs from Wikidata and Google cover up to several hundred nodes. 

Ostendorff \textit{et al.} \shortcite{Ostendorff:2019} enrich BERT with the author information via Wikidata to improve book classification. However, they struggle to select the relevant properties to traverse the Wikidata knowledge graph. Instead, they use pre-trained graph embeddings as author representations, trained on the full Wikidata graph. In contrast, we use LDA to filter the properties that match the input-specific context. Their model is not applicable for argument mining, which is why we can not compare to them.

While the above approaches build upon pre-trained language models, other approaches enhance language models, more specifically BERT, with external knowledge \cite{Guan:2020, He:2019, Levine:2019, Zhang:2019, Xiong:2019, Peters:2019, Wang:2020, Lauscher:2019}. BERT, as the current state-of-the-art model, is pre-trained on two objectives simultaneously: masked language modeling and next sentence prediction.



\section{Method}

Our goal is to extract relevant paths from the knowledge graph that may help in the downstream argument mining task. The paths connect entities from the given topic and sentence. The challenge is to do this efficiently despite the huge number of entities and properties present in the knowledge graph. Additionally, the path should be focused on a specific topic to ensure that it is a relevant path and not a connection based on random hops.
We train LDA on the Wikipedia articles of each entity to choose more suitable properties for growing the knowledge graph. Using TF-IDF, we filter the highest scoring properties for the topics given by the model. 
Avoiding areas in the graph that are not within the given sentence context, we map each token of an entity to a specific word vector and check whether its cosine similarity with the sentence vector passes a certain threshold. 
Additionally, we build a second knowledge graph based on unstructured data from Google search for a given topic. Annotating the documents via OpenIE, we combine the two graphs to tackle the incompleteness of Wikidata. 

The whole process is visualized in Figure \ref{gfx:total}, the topic and sentence are mapped to the Wikidata entities via the Wikifier \cite{Brank:2017}, an online entity linker, and the local knowledge graph built from Wikidata. Wikidata, an open structured knowledge graph, stores information in the form of (entity, property, entity)-tuples. This allows us to iteratively build a local graph with Breadth-First Search (BFS) and finally extract evidence paths. We focus on two kinds of information: (1) paths between the topic and the sentence, signaling topic relevance, and (2) paths between tokens within a sentence to gather additional world knowledge (``evidence'') for it.

\begin{figure}[t]
    \centering
    \includegraphics[width=1.0\columnwidth,trim=9 10 5 10,clip,]{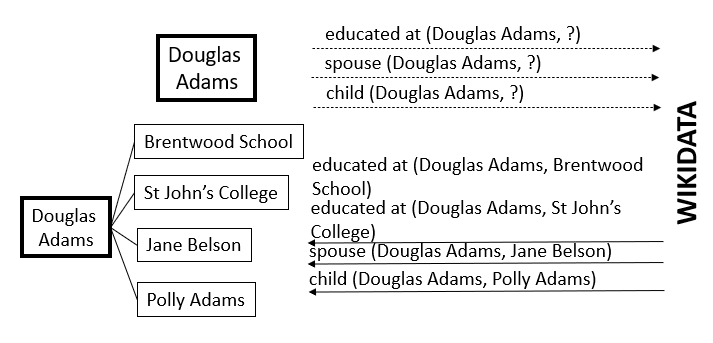}
    \caption{Expansion of the node ``Douglas Adams" via a SPARQL query (dashed arrow) per entity per property.}
    \label{gfx:sparql_old}
\end{figure}

\subsection{Main Core}

An argument is a combination of a topic and a sentence holding evidence towards this topic, which are given as input to the method. As shown in Figure \ref{gfx:total}, 
these are the inputs to the Wikifier, which annotates the input document with relevant Wikipedia concepts via a PageRank-based method. The output is a JSON document containing a list of annotated Wikipedia concepts along with their corresponding Wikidata entities. As an input, we consider the top $k_s$ concepts found in the sentence and the top $k_t$ concepts found in the topic, each concept representing a specific token.

Our main source of knowledge is the collaboratively constructed knowledge-base Wikidata. Wikidata is free, multilingual, and its broad community curation ensures a high data quality with currently more than 88 million items $E$ and 5519 properties $R$. Beyond a label (e.g., ``Douglas Adams'') and an identifier (e.g., ``Q42''), each item covers a set of statements $S$ linking to other items. A statement, expressing semantic or ontological connections, can be described as a binary relation between entities (e.g., P69(Q42, Q691283) represents the fact that Douglas Adams (Q42) got educated at (P69) St John's College (Q691283)). Formally, we describe Wikidata as a graph $G:=(E,R,S)$ with $S=\{r(e_1,e_2)| r\in R, e_1,e_2\in E\}$. Each entity (e.g. ``Douglas Adams (Q42)'') is connected to several other entities via specific properties (e.g. ``educated at (P69)''), creating a relation-based knowledge structure. Therefore, it can be represented as a list of binary relations (e.g. ``educated at''(``Douglas Adams'', ``St John's College'')). Graphically, this list can also be seen as a list of triples with each entity being a node and each relation being an edge connecting these two nodes. Starting from this graph, in every of the $n_d$ iterations, it queries Wikidata via SPARQL-queries for a list of entities connected to one of the unseen nodes in the graph via one of the properties of interest. 
Figure~\ref{gfx:sparql_old} represents a sample Wikidata query with its returned list of object-entities. Repeating this BFS-like expansion in every iteration, a knowledge graph representing the Wikidata-based concepts and their relations regarding the given topic and sentence is extracted. To cover both sentence-sensibility and sentence-topic-coherence, we extract two kinds of paths from the graph using NetworkX\footnote{\url{https://networkx.github.io/}}: (1) the shortest path connecting one topic-concept with one sentence-concept and (2) the shortest path connecting two sentence-concepts. These paths serve as evidence to help classify whether the given sentence is an argument to the given topic.

\subsection{Property Selection with Latent Dirichlet Allocation}

Building a knowledge graph over all the existing properties per node is infeasible as: (1) In the worst case, at depth $D$ the graph already has $\sum_{d=0}^D{E_0P^D}$ nodes and $\sum_{d=1}^D{E_0P^D}$ edges with $E_0$ being the number of nodes that we started with 
and $P$ the number of properties. 
With the BFS runtime of a graph being $\mathcal{O}(|edges|+|vertices|)$, the search is not feasible for $P$=5519. (2) Most of the properties do not help gain relevant information on the given concepts, they instead drastically increase the noise. (3) Building the graph based on a small fixed set of properties leaves out most of the available information. (4) Alternatively, taking a predetermined set of properties (e.g., the 50 most frequent ones) lacks coherence to the given context. 
Considering these challenges, we select relevant properties to the context of a specific sentence and topic dynamically by latent Dirichlet allocation. 

\SetKwInput{KwInput}{Input}
\SetKwInput{KwOutput}{Output}

\begin{algorithm2e}[tb]
\caption{PROPERTIES\_PER\_ENTITIES}\label{alg:lda}
\DontPrintSemicolon
\KwInput{entities $E$, tfidf-threshold $t_t$, count-threshold $t_c$, num-topics $n_t$, num-properties $n_p$, property-descriptions $P$}

\nl	$D \leftarrow \emptyset$ \;
\nl \ForAll{$e\in E$}{%
\nl $D\leftarrow D\cup$Wikipedia article of $e$\;
}
\nl $L \leftarrow$ LDA($D, n_t$) \;
\nl $T \leftarrow$ top-topics($L$)\;
\nl $W \leftarrow \emptyset$ \;
\nl \ForAll{$t \in T$}{%
\nl \ForAll{$w \in t$}{%
\nl \If{$w\notin W \textbf{ and } w\notin$ stopwords}{
\nl $W \leftarrow W\cup w$ \;
}}}
\nl $F \leftarrow$ \textit{RANK\_BY\_TFIDF}($W,t_t$) \;
\nl $\tilde{P} \leftarrow \emptyset$\;
\nl \ForAll{$w \in F$}{
\nl \ForAll{$p \in P$}{
\nl \If{$p \notin \tilde{P} \textbf{ and } w \in p.info \textbf{ and } p.count>t_c$}{
\nl $\tilde{P} \leftarrow \tilde{P}\cup p$ \;
}}}
\nl \Return{SELECT\_FREQUENT($\tilde{P},n_p$)}
\end{algorithm2e}

As shown in Figure \ref{gfx:total}, the blue box covers three steps: (1) retrieving Wikipedia articles on the given entities, (2) training an LDA model on the articles to discover properties related to this specific context, and (3) embedding the properties into the main core.
While the first and third steps are lightweight in terms of preprocessing effort, in the second step, we need to find a way to compare the properties of Wikidata to the LDA output. Fortunately, Wikidata provides a list of all properties with a description. This information serves as an interface between the properties and the words representing the topics given by LDA (listed as input in Algorithm~\ref{alg:lda} under ``property\_descriptions'').
We derive the list of properties given a list of entities in five steps (Algorithm~\ref{alg:lda}): 
\begin{enumerate}
    \item Load the corresponding Wikipedia articles using the Wikipedia-API to find the given entities' articles and extract the relevant texts (lines~1-3). 
    \item Train an LDA model to extract the most relevant topics for this specific context (line~4). Preprocessing steps include removing stopwords, punctuations, and the numeric expressions from the articles after converting them to lowercase, and finally, applying tokenization and lemmatization. We build a corpus based on the preprocessed text, which we apply LDA on. Then, we extract the $n_t$ best topics represented as a batch of words each (line~5).
    \item Rank the words representing the topics by their relevance to the properties measured via TF-IDF (lines~6-11). In order to retrieve the relevant yet indispensable words, we consider the property descriptions as a batch of documents. We build a matrix \(M\) with each cell \(m_{i,j}\) being the TF-IDF of the \(i\)-th word and the \(j\)-th document and rank the words by their cumulative score 
    
    \begin{equation*}
        S_i=\sum_{j}{m_{i,j}},
    \end{equation*}
    proceeding with only those achieving a given threshold.
    \item Extract top-related properties by searching the property descriptions for the top-ranked words (lines~12-16). We consider every property with at least one word appearing at least once. Ranked by their number of appearances (line~17), we return the $n_p$ most frequent properties.
\end{enumerate}

\subsection{Knowledge Retrieval with Dynamic BFS and Word
Embedding}
Only one node connection out of the context can result in a noisy path when building a graph dynamically. 
For instance, Figure~\ref{gfx:problem_input} shows one possible path connecting the concepts \underline{offices} and \underline{times} within the sentence.
This noisy path can only be found since the entity \textbf{\textit{spacetime (Q133327)}} connects the \textbf{offices}-related concepts and the \textit{times}-related concepts, even though, they do not fit into this sentence-specific context. 
Furthermore, if concepts are relevant with respect to more than one topic, they build a knowledge graph with a set of subgraphs, each covering the knowledge of one particular topic. Traversing such diverse graph is not optimal, instead, deeper exploration of one relevant subgraph could yield important information. 
We address these challenges by only expanding child nodes if the GloVe word embedding~\cite{Pennington:2014} of every node (entity vector) belongs to the context of the input sentence (sentence vector) with respect to a cosine similarity threshold. 
GloVe primarily performs well on word analogy, word similarity, and named entity recognition tasks, which makes it suitable to our setting. 
We lower the computational complexity by not querying once per (entity, relation)-pair but per entity, enabling a theoretical speedup of the size of properties.

\begin{algorithm2e}[tb]
\caption{DYNAMIC\_BFS} \label{alg:bfs}
\DontPrintSemicolon
\KwInput{sentence $s$, concepts $C$, entities $E$, embedding-model $GV$, cosine-threshold $t_{cos}$, max-nodes $n_n$, max-depth $n_d$}

\nl $G \leftarrow$ MultiDiGraph() \;
\nl \For{$i=0 \hdots \#E-1$}{
\nl $G\leftarrow G\cup edge$($C_i, $``WIKIFIERED''$, E_i$)\;
}
\nl $v_s\leftarrow$ avg($\{GV$($t$)$|t\in$ tokenize(lower($s$))$\}$)\;
\nl $P\leftarrow$ \textit{PROPERTIES\_PER\_ENTITIES}($E$)\;
\nl $d\leftarrow 0, E_{visited}\leftarrow \emptyset, E_{tovisit} \leftarrow E$\;
\nl \While{$E_{tovisit}\neq\emptyset \textbf{ and } d<n_d \textbf{ and } \# E_{visited} < n_n$}{
\nl $d\leftarrow d+1$\;
\nl \ForAll{$e\in E_{tovisit}$}{
\nl $E_{tovisit}\leftarrow E_{tovisit}\setminus e$\;
\nl \If{$e\notin E_{visited}$}{
\nl $E_{visited}\leftarrow E_{visited}\cup \{e\}$\;
\nl \ForAll{($e, p, e_{new}$) $\in$ QUERY($e,P$)}{
\nl \If{$e_{new}\notin E_{visited}\textbf{ and } e_{new}\notin E_{tovisit}\textbf{ and }$ COSINE\_SIMILARITY($GV, v_s, e_{new}, t_{cos}$)}{
\nl $E_{tovisit}\leftarrow E_{tovisitd}\cup \{e_{new}\}$\;
\nl $G\leftarrow G\cup edge$($e, p, e_{new}$)\;
}}}}}
\nl \Return{$G$}\;
\end{algorithm2e}

\begin{figure}[tb]
	\centering
  \includegraphics[width=0.7\columnwidth]{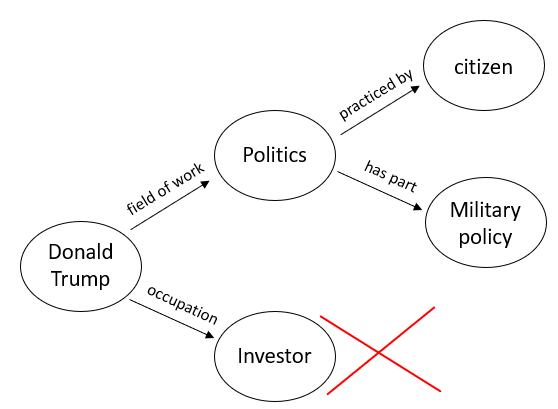}
	\caption{Visualization of how word embeddings make the graph more sparse.}
	\label{gfx:wordvec}
\end{figure}

We map each word to its GloVe embedding and propose an augmentation method to any BFS-based graph construction in Algorithm~\ref{alg:bfs}. We initialize a directional graph and allow multiple edges between two nodes via NetworkX\footnote{\url{https://networkx.github.io/}}. For every concept in the given topic or sentence which is annotated by an entity via the Wikifier, we add two nodes and an edge to the graph to represent their ``WIKIFIERED''-relation (line 2--3). These concept- and entity-nodes build the basis of our graph. 
Before further growth of the graph, we calculate the sentence vector $v_s$ as the average of the word vectors contained in the sentence (line 4) via the GloVe model, $GV$. %
Using Algorithm~\ref{alg:lda}, we receive a list of properties to build the graph  
(line 5). Lines 6-16 present a modified version of the standard BFS algorithm by querying Wikidata (line 13) and pruning the graph (line 14). 
To improve the query time of the increasing number of properties, we adapt the SPARQL-queries by submitting a Wikidata query per entity instead of one query per (entity, property)-pair. The returned list of (subject, predicate, object) statements contains potential new object-entities. However, before adding a new entity $e_{new}$ to the graph, we prove it to fulfill one of the following conditions:
\begin{linenomath}
\begin{align*}
    \max\{cos(v_s, v_t) | t\in e_{new}\} &> t_{cos}\\
    \max\{cos(v_s, v_t) | t\in e_{new}\} &= -1, 
\end{align*}
\end{linenomath}
where $t_{cos}$ is a threshold and $v_t$ is the $GV$-vectors of the lowercase tokens in $e_{new}$. We thereby assure at least one of the tokens in the new entity stays in the context of the given sentence. Being equivalent to -1 means that $GV$ covers no token in the entity. We make sure that particular entities do not vanish. Therefore, we only exclude entities with a maximum cosine being strictly greater than -1 but less than $t_{cos}$. Figure~\ref{gfx:wordvec} visualizes the exploration of Wikidata into the sentence-specific context for a sample input sentence ``Trump uses the military to prove his manhood''.
Calculating the cosine similarities yields:
\begin{linenomath}
\begin{align*}
    cos(v_s,GV(\text{``Trump''}) &>t_{cos}\\
    cos(v_s, GV(\text{``Politics''})) &>t_{cos}\\
    -1<cos(v_s, GV(\text{``Investor''})) &<t_{cos}
\end{align*}
\end{linenomath}
Therefore, the sentence-specific relevant nodes ``Donald Trump'' and ``Politics'' expand, the node ``Investor'' does not.

\begin{algorithm2e}[tb]
\caption{ENRICH} \label{alg:openie}
\DontPrintSemicolon
\KwInput{graph $G=(V_G,E_G)$, topic $t$, max-urls $n_u$, max-annotations $n_a$, max-chars $n_{c_{max}}$, min-chars $n_{c_{min}}$}
\nl $U \leftarrow$ \textit{PAGE\_RANK}($t, n_u$)\;
\nl $S \leftarrow \emptyset$\;
\nl \ForAll{$u\in U$}{
\nl \ForAll{$s\in u.\mbox{text}$}{
\nl $S\leftarrow S\cup s$\;
}}
\nl $S\leftarrow$ \textit{SORT\_BY\_SIZE}($S$)\;
\nl $\mbox{corpus} \leftarrow \textit{EXHAUST}(S, n_{c_{max}}, n_{c_{min}})$\;
\nl $A \leftarrow \textit{OPENIE\_ANNOTATE}(\mbox{corpus})$\;
\nl \For{$i=0\rightarrow n_a$}{
\nl $(\mbox{subject}, \mbox{predicate}, \mbox{object}) \leftarrow A_i$\;
\nl $G \leftarrow G+ edge(\mbox{subject}, \mbox{predicate}, \mbox{object})$\;
\nl \ForAll{$v\in V_G$}{
\nl \If{MATCH$(v,\mbox{subject})$}{
\nl $G \leftarrow G\cup edge(v, \mbox{``OPENIED''}, \mbox{subject})$\;
}
\nl \If{MATCH$(v,\mbox{object})$}{
\nl $G \leftarrow G\cup edge(v, \mbox{``OPENIED''}, \mbox{object})$\;
}}}
\nl\Return{$G$}\;
\end{algorithm2e}

\subsection{Enriching the KB-graph with OpenIE}

\begin{figure}[tb]
	\centering
  \includegraphics[width=0.7\columnwidth]{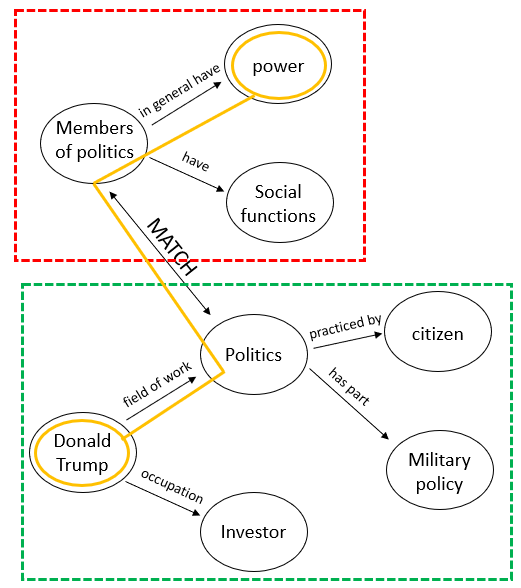}
	\caption{
	Given the input sentence, ``\underline{Donald Trump} has a lot of \underline{power}.", the structured data-driven knowledge graph (green box) and the unstructured data from Google (red box) are combined via a ``MATCH'' function. One can discover the yellow path, which would not have been possible with only one source of knowledge.}
	\label{gfx:openie}
\end{figure}

Recent studies proved the importance of additional unstructured data to compensate for the incompleteness of structured knowledge-bases like Wikidata \cite{Potash:2017, Lv:2020, Clark:2019}. However, most approaches struggle with picking the relevant data and/or handling the noise. 
Instead of using either hand-picked or randomly chosen articles \cite{Potash:2017}, we consider top-ranked articles based on PageRank found in the Google searches for a given topic. Using OpenIE \cite{Manning:2014}, we build a second knowledge graph and combine them gradually to handle the noise.
We enrich the knowledge graph with unstructured data in 4 steps (Algorithm~\ref{alg:openie}):
\begin{enumerate}
    \item By searching Google for a given topic, 
    we receive a list of websites ranked by Google's PageRank algorithm (line 1). Considering the top $n_u$ websites, we extract each sentence with at least three words (line 5) -- the minimum to let OpenIE build a triple from. For various reasons (runtime, storage, and OpenIE limits), we rank the sentences by their total character length (line 6) because we assume that longer sentences tend to be more relevant.
    \item Extract (subject, predicate, object)-triples via Stanford's information extraction framework, OpenIE, which takes the corpus as input (line 8). These triples, representing statements, are the basis of our second knowledge graph after removing duplicates.
    \item Build the graph based on the first $n_a$ statements. If for example, the sentence ``In general members of politics have power'' is annotated with the triple (``Members of politics'', ``in general have'', ``power''), the subject and object become nodes while the relation converts to an edge between them (line 11). 
    \item Combine the two knowledge graphs $G_s=(V_s,E_s)$ and $G_u=(V_u,E_u)$ by summing up the set of edges $\mathcal{V}$:
\begin{linenomath}    
\begin{equation*}
\begin{split}
   \mathcal{V}= \{(u,v)\in V_s\times V_u | \exists t: t\in u, t\in v, t\notin \mbox{stopwords}, \\t\notin \mbox{numerics}, |e|>2\}, 
   \end{split}
\end{equation*}
\end{linenomath}
with each node being a set of tokenized and lemmatized tokens (lines 12-16). Figure~\ref{gfx:openie} illustrates this procedure for a sample sentence.
\end{enumerate}

\section{Argument Identification}
We model the task of argument identification as a binary classification problem. Since we explicitly do not focus on the stance identification of arguments, we do not differentiate between pros and cons; hence, we only consider two labels, argument or no-argument. We use the following models for the prediction task:

\noindent\textbf{Baseline:} As a baseline, we use a simple BiLSTM from Stab \textit{et al.} \shortcite{Stab:2018} that only receives the sentence as an input.

\noindent\textbf{Baseline+X:} To incorporate the collected evidence paths for each sentence, we use two BiLSTMs, and an attention mechanism. One BiLSTM is used to flatten each path to a single vector, and the other BiLSTM encodes the sentence. Attention is used to generate a single vector from an embedded token of a sentence and all path vectors connected to that.

We define a path as a sequence $e_0, p_0, e_1, p_1, ..., e_n$, consisting of entities, $e$, and predicates, $p$, embedded with pre-trained knowledge graph embeddings. We use a BiLSTM (shared among all paths) to reduce each path to a single vector $q$ (last hidden state of that BiLSTM). All paths $(q_1, q_2, ..., q_m)$ for an embedded token $v$ of an input sentence are then used in an attention mechanism to gain a final vector $u$. 
The attention mechanism is defined as follows \cite{Hermannteaching}:
\begin{linenomath}
\begin{gather}
\hspace*{-3mm}\pmb{m}_{i} = \tanh(\pmb{W}_{q}\pmb{q}_{i} + \pmb{W}_{v}\pmb{v})\label{eq:outer_att_1},\\
\hspace*{-3mm}f_\mathit{attention}(\pmb{q}_{i}, \pmb{v}) = \frac{\exp(\pmb{w}^T_{m}\pmb{m}_{i})}{\sum_{i}\exp(\pmb{w}^T_{m}\pmb{m}_{i})},\label{eq:outer_att_2}
\end{gather}
\end{linenomath}
where $\pmb{W}_q$, $\pmb{W}_v$, and $\pmb{w}_m$ are trainable weight parameters, and $q_i$ denotes the $i$-th path of a token $v$. With the attention weights, we compute the final, weighted path vector $\pmb{u}$ as: 
\begin{linenomath}
\begin{gather}
\alpha_i \propto f_\mathit{attention}(\pmb{q}_{i}, \pmb{v}),\label{eq:outer_att_3}\\
\pmb{u} = \sum_{i=1}^{n}\pmb{q}_{i}\alpha_i.\label{eq:outer_att_5}
\end{gather}
\end{linenomath}
Vector $u$ is then concatenated with the token embedding $v$ and passed into the sentence encoder BiLSTM. The procedure is repeated for all tokens of the input sentence. Finally, we take the last hidden state of the sentence encoder BiLSTM as the input for a classification layer with two neurons and use a Softmax activation function to determine the class label.

\section{Experimental Evaluations}

\setlength{\tabcolsep}{6pt}
\begin{table}[t]
    \centering
    \small
    \begin{tabular}{lrrrr}
        \hline
        topic & sentences & neutral & pro & con\\
        \hline
        abortion & 3,929 & 2,427 & 680 & 822 \\
        cloning & 3,039 & 1,494 & 706 & 839 \\
        death penalty & 3,651 & 2,083 & 457 & 1,111 \\
        gun control & 3,341 & 1,889 & 787 & 665 \\
        marijuana legalization & 2,475 & 1,262 & 587 & 626 \\
        minimum wage & 2,473 & 1,346 & 576 & 551 \\
        nuclear energy & 2,576 & 2,118 & 606 & 852 \\
        school uniforms & 3,008 & 1,734 & 545 & 729 \\
        \hline
        total & 25,492 & 14,353 & 4,944 & 6,195 \\
        \hline
    \end{tabular}
    \caption{Characteristics of the argument mining test corpus.}
    \label{tab:testcorpus}
\end{table}

We compare our framework to the baseline and partial improvements on eight datasets of the UKP Sentential Argument Mining Corpus (Table \ref{tab:testcorpus}) \cite{Gurevych:2018}. We run our experiments on a single device with a Intel Core i5-10210U CPU and a 16GB RAM. Each dataset holds a collection of sentences on one specific topic, out of which we test on 200 each. The instances cover the following controversial topics: Abortion, cloning, death penalty, gun control, marijuana legalization, minimum wage, nuclear energy, and school uniforms. Each instance either supports or opposes a topic, or is off-topic. We train the knowledge graph embeddings for the classifier on a Wikidata dump with openKE \cite{Han:2018}. For the BiLSTM we used the following hyperparameters: $dropout=0.7,~ lstm\_size=64,~ batch\_size=16,~ learning\_rate=0.001,~ epochs=10,~ attention\_size=50,~ max\_paths=10,~ max\_path\_len=15$. 
We define the main core and upgrade it with the previously discussed methods for further investigations:

\setlength{\tabcolsep}{5pt}
\begin{table*}[t]
    \centering
    \small
    \begin{tabular}{lccccccccc}
        \hline
        model & \makecell{sents with\\sen$\rightarrow$sen} & \makecell{sents with\\sen$\rightarrow$top} & \makecell{avg\\ \#sen$\rightarrow$sen} & \makecell{avg\\ \#sen$\rightarrow$top} & \makecell{avg\\ \#hops} & \makecell{avg\\path len} & \makecell{avg\\runtime} & acc & F1\\
        \hline
        Baseline &&&&&&&&& .6069 $\pm$ .0074\\
        +WD & .3316 & .3816 & 0.14 & 0.86 & \textbf{7.29} & 1.62 & 30.16 & .7817 $\pm$ .0095 & .6359 $\pm$ .0162\\
        +WD+LDA & .4400 & .5125 & 1.15 & 1.39 & 5.15 & 3.77 & 276.21 & .7961 $\pm$ .0075 & .6587 $\pm$ .0227\\
        +WD+LDA+GV & .4888 & .5462 & 1.47 & 1.57 & 6.44 & 4.45 & 301.75 & .8163 $\pm$ .0152 & .6227 $\pm$ .0368\\
        +WD+LDA+GV+OIE & \textbf{.8038} & \textbf{.7012} & \textbf{6.89} & \textbf{9.79} & 6.66 & \textbf{7.17} & 334.52 & \textbf{.8504} $\pm$ .0104 & \textbf{.6785} $\pm$ .0214\\
        \hline
    \end{tabular}
    \caption{Results for each model on the eight test sets over ten seeds. Paths between entites within the sentence are referenced as ``sen$\rightarrow$sen'' and connecting the sentence with the topic as ``sen$\rightarrow$top''. Bold numbers indicate the highest score in the column.}
    \label{tab:results}
\end{table*}

\noindent\textbf{+WD} only takes the two most frequent Wikidata properties (``subclass of'', ``instance of'') into account to build the knowledge graph in maximum $n_d=10$ iterations, focusing up to $k_s=5$ concepts in the sentence and $k_t=2$ in the topic, as our preliminary experiments have shown to return good results.

\noindent\textbf{+WD+LDA} uses the top $n_p=50$ properties selected via an LDA model of 200 iterations. We set the number of topics to $n_t=5$, the TF-IDF threshold to $t_t=2.5$, and the minimum occurrence of a property on Wikidata to $t_c=1000$. 

\noindent\textbf{+WD+LDA+GloVe} includes word embeddings to avoid out-of-context parts in the graph. Focusing on low complexity, we use 50-dimensional vectors pre-trained on Wikipedia 2014 and Gigaword 5\footnote{\url{https://catalog.ldc.upenn.edu/LDC2011T07}}.
We choose a cosine similarity cutoff $t_{cos}=0.4$, which sample-based error analysis has shown a good performance. Thus, we build a graph covering up to $n_n=600$ nodes.

\noindent\textbf{+WD+LDA+GloVe+OpenIE} enriches the aforementioned graph with unstructured data via OpenIE. From the first $n_u=3$ websites we process between $n_{cmin}=70000$ and $n_{cmax}=99999$ characters to let OpenIE extract up to $n_a=600$ annotations.


As shown in Table \ref{tab:results}, our main core including evidence paths via Wikidata (+WD) improves on the baseline by 5\% on F1 score (from 0.6069 to 0.6359). Selecting fifty properties, based on LDA, instead of the two most frequent ones not only improves on accuracy and F1 score but also fits best to the given context. In terms of extracted paths, it increases the number of sentences with at least one path connecting sentence entities by over 10\%, and with at least one path connecting the sentence and the topic by over 13\%. While the number of paths and the average path length increase, the average number of processed hops, which measures the overall searchable depth, decreases from 7.29 to 5.15. This is the result of the edges per node (properties per entity) going from 2 to 50. Tackling that issue, sparsing the graph using word embeddings successfully increases the hop count to 6.44. Breaking 81\% on accuracy, it extracts at least one path within the sentence in 48 out of 100 cases and a connecting sentence and topic in 54 out of 100 cases. Enriching the graph with unstructured knowledge not only significantly increases the quantitative measuring numbers regarding the paths, but also leads to outstanding 85\% accuracy and 67\% F1-score, improving on the baseline by 7\%. 


\subsection{Error Analysis}

\setlength{\tabcolsep}{7pt}
\begin{table}[tb]
    \centering
    \begin{tabular}{lll}
        \hline
        Topic & Sentence & Label \\
        \hline
        Abortion & \makecell[l]{There is no third \\possibility.} & NoArgument\\
        Nuclear energy & We hate spam too! & NoArgument\\
        \hline
    \end{tabular}
    \caption{Sentences for which irrelevant paths are found.}
    \label{tab:error}
\end{table}

To investigate under what conditions our framework finds irrelevant paths, we conduct a qualitative error analysis. We randomly selected error paths which were found by our framework. The reasons are mainly (1) questionable entity linking by Wikifier, (2) falsely connecting nodes from the two graphs, and (3) noisy data. The first example in Table \ref{tab:error} generates the following path:
\begin{lstlisting}[mathescape, frame=none, breaklines]
  $\textbf{There is}\xrightarrow[]{WIKIFIERED}\textit{English grammar (Q560583)}$
  $\xrightarrow[P361]{part~of}\textit{English (Q1860)}\xrightarrow[P17]{country}$
  $\textit{United States of America (Q30)}\xrightarrow[]{MATCH}\textit{state}$
  $\xrightarrow[]{considered}\textit{abortion}\xrightarrow[]{MATCH}\textbf{abortion}.$
\end{lstlisting}
Due to the Wikifier mapping ``There is'' to the Wikidata entity ``English grammar'' this path falsely connects the sentence to the topic. This problem can be mitigated by specifying the Wikifier parameters less error-prone. The second example in Table \ref{tab:error} produces the path:
\begin{lstlisting}[mathescape, frame=none, breaklines]
  $\textbf{hate}\xrightarrow[]{WIKIFIERED}\textit{Hate speech (Q653347)}$
  $\xrightarrow[P5008]{on~focus~list~of~Wikimedia~project}$
  $\textit{WikiForHumanRights (Q78499962)}\xrightarrow[P31]{instance of}$
  $\textit{human rights (Q8458)}\xrightarrow[]{MATCH}\textit{humanity}\xrightarrow[]{has}$
  $\textit{energy~needs}\xrightarrow[]{MATCH}\textbf{energy}.$
\end{lstlisting}
Although every edge in the path makes some sense, the total connection of ``hate'' and ``energy'' via the context of human rights is rather questionable. This problem can be mitigated by applying further cleaning strategies on the used knowledge.

\subsection{Case Study}
The example in Figure \ref{gfx:problem_input} demonstrates the power of our method. Calculating the sentence vector $v_s$ and looking up the entity vectors yields:
\begin{linenomath}
\begin{align*}
    &cos(v_s,GV(\text{``office''}))&\approx &0.7007,\\
    &cos(v_s,GV(\text{``room''}))&\approx &0.7195,\\
    &cos(v_s,GV(\text{``location''}))&\approx &0.6469,\\
    &cos(v_s,GV(\text{``space''}))&\approx &0.6210,\\
    &cos(v_s,GV(\text{``spacetime''}))&\approx &-0.0365,\\
    &cos(v_s,GV(\text{``time''}))&\approx &0.8891.
\end{align*}
\end{linenomath}
Since the concepts of office and time are relevant to this specific sentence, it makes sense that the vectors of corresponding entities have a cosine similarity close to 1. Whereas the entity ``spacetime'' might be related to both space and time, it is not relevant to this sentence. Therefore, cutting at the threshold of $t_{cos}=0.4$ utilizes this fact and prevents finding this path.

\section{Conclusion}

We presented a novel argument mining framework, which improves knowledge extraction using structured and unstructured data. We overcome the problem of exponential growth of the knowledge graph needed for path extraction using two key ideas: topic modeling to improve the selection of relevant properties, and word embedding to ensure topical consistency, which leads to a sparser knowledge graph. 
In comparison to existing methods, we can process a much larger amount of data and take more possible properties into consideration. This allows us to discover more relevant paths. Our results show an average performance of 85\% for the accuracy and 67\% for the F1 measure on the UKP sentential argument mining corpus. This may contribute to future work that combines argument mining with discrete knowledge in the form of knowledge graphs.

\section{Acknowledgements}
Benjamin Schiller was supported by the German Research Foundation within the project ``Open Argument Mining'' (GU 798/25-1), associated with the Priority Program ``Robust Argumentation Machines (RATIO)'' (SPP-1999).

\bibliography{bibrefs}

\end{document}